\documentclass[twocolumn,showpacs,preprintnumbers,superscriptaddress]{revtex4}
\usepackage[T1]{fontenc}     
\usepackage{amsfonts}
\usepackage{amsmath,amsbsy,amssymb,graphicx}              
\usepackage{times} 
   
\begin{document}                          
     
\title{Spin  Supercurrent in the Canted Antiferromagnetic Phase}
\author{Yusuke~Hama}
\affiliation{Department of Physics, The University of Tokyo, Tokyo 113-0033,  Japan}
\affiliation{Theoretical Research Division, Nishina Center, RIKEN, Wako 351-0198, Japan}
\author{George~Tsitsishvili}
\affiliation{Department of Physics, Tbilisi State University, Tbilisi 0128, Georgia}
\author{Zyun~F.~Ezawa}
\affiliation{Advanced Meson Science Laborary, Nishina Center, RIKEN, Wako 351-0198, Japan}
\date{\today}
\preprint{RIKEN-QHP-51}      
\begin{abstract}
The spin and layer (pseudospin) degrees of freedom are entangled coherently
in the canted antiferromagnetic phase of the bilayer quantum Hall system at
the filling factor $\nu =2$. There emerges a complex Goldstone mode
describing such a combined degree of freedom. In the zero
tunneling-interaction limit ($\Delta_{\text{SAS}}\rightarrow 0$), its phase
field provokes a supercurrent carrying both spin and charge within
each layer. The Hall resistance is predicted to become anomalous precisely
as in the $\nu =1$ bilayer system in the counterflow and drag experiments.
Furthermore, it is shown that the total current flowing in the bilayer system
is a supercurrent carrying solely spins in the counterflow geometry.
It is intriguing that all these phenomena occur only in imbalanced bilayer
systems.
\end{abstract}
 
\maketitle

\section{Introduction}            
Physics of the bilayer quantum Hall (QH) system is enormously rich owing to
the intralayer and interlayer phase coherence controlled by the interplay
between the spin and the layer (pseudospin) degrees of freedom\cite{Ezawa:2008ae,S. Das Sarma1}. 
At the filling factor $\nu =1$ there arises a
unique phase, the spin-ferromagnet and pseudospin-ferromagnet phase, which
has well been studied both theoretically and experimentally. One of the most
intriguing phenomena is the Josephson-like tunneling between the two layers
predicted in Refs.\cite{EisensteinMacDonald,Ezawa:1992,Wen}, whose first experimental indication
was obtained in Ref.\cite{Spielman1}. Other examples are the anomalous
behavior of the Hall resistance reported in counterflow experiments\cite{Kellog1,Tutuc} 
and in drag experiments\cite{Kellog2}. They are triggered by 
the supercurrent within each layer\cite{Ezawa:2007nj}. Quite
recently, careful experiments \cite{Tiemann} were performed to explore the
condition for the tunneling current to be dissipationless. These phenomena
are driven by the Goldstone mode describing an interlayer phase coherence.   
It exhibits the linear dispersion relation
in the zero tunneling-interaction limit ($\Delta_{\text{SAS}}\rightarrow 0$).

On the other hand, at $\nu =2$ the bilayer QH system has three phases, the
spin-ferromagnet and pseudospin-singlet phase, the spin-singlet and
pseudospin ferromagnet phase, and a canted antiferromagnetic phase\cite{S.
Das Sarma2,MacDonald,Pellegrini1,khrapai1,Sawada4} (abridged as the CAF phase),
depending on the relative strength between the Zeeman energy 
$\Delta_{\text{Z}}$ and the tunneling energy $\Delta_{\text{SAS}}$. 
The pattern of the symmetry breaking is 
SU(4)$\rightarrow $U(1)$\otimes $SU(2)$\otimes $SU(2),   
associated with which there appear four complex Goldstone
modes\cite{Hasebe}. 
A part of them has been studied in Refs.\cite{nu2goldstone,Hasebe}.   
We have recently analyzed the full details of these
Goldstone modes in each phase\cite{yhama}. 
The CAF phase is the most interesting, where 
the spins are canted coherently and making antiferromagnetic correlations 
between the two layers. Moreover, 
one of the Goldstone modes becomes gapless and has a
linear dispersion relation\cite{yhama} as 
$\Delta_{\text{SAS}}\rightarrow 0$. It is an urgent and intriguing
problem what kind of phase coherence this Goldstone mode develops.

In this paper, we show that it is the entangled spin-pseudospin phase
coherence, and we explore associated phase coherent phenomena. We employ the
Grassmannian formalism\cite{Hasebe}, where the basic field is the
Grassmannian field consisting of two complex-projective ($\text{CP}^{3}$)
fields. The CP$^{3}$ field emerges when composite bosons undergo
Bose-Einstein condensation\cite{Ezawa:2008ae}. The formalism provides us
with a clear physical picture of the spin-pseudospin phase coherence in the
CAF phase. Furthermore, it enables us to analyze nonperturbative phase
coherent phenomena, where the phase field $\vartheta (\mathbf{x})$ is
essentially classical and may become very large. We show that the 
supercurrent flows within the layer when there is inhomogeneity in $\vartheta (\mathbf{x})$. 
This is precisely the same as in the $\nu =1$
bilayer QH system. Indeed, the supercurrent leads to the same formula\cite{Ezawa:2007nj} 
of the anomalous Hall resistivity for the counterflow and
drag geometries as the one at $\nu =1$. What is remarkable is that the total
current flowing the bilayer system is a supercurrent carrying solely
spins and not charges in the counterflow geometry. We note that the supercurrent
flows both in the balanced and imbalanced systems at $\nu =1$ but only in
imbalanced systems at $\nu =2$.

\section{Grassmannian field and  entangled spin-pseudospin phase coherence }   
In the bilayer system an electron has two types of indices, the spin index 
$(\uparrow ,\downarrow )$ and the layer index $(\text{f},\text{b})$. They can
be incorporated into 4 types of isospin index 
$\alpha =$ f$\uparrow $,f$\downarrow $,b$\uparrow $,b$\downarrow $. 
The electron field $\psi_{\alpha}(\boldsymbol{x})$ 
has four components, and the bilayer system possesses the
underlying algebra SU(4) with the subalgebra 
$\text{SU}_{\text{spin}}$(2)$\otimes \text{SU}_{\text{ppin}}$(2). 
We denote the three generators of the 
$\text{SU}_{\text{spin}}$(2) by $\tau_{a}^{\text{spin}}$, and those of 
$\text{SU}_{\text{ppin}}$(2) by $\tau_{a}^{\text{ppin}}$. 
There are remaining nine generators $\tau_{a}^{\text{spin}}\tau_{b}^{\text{ppin}}$,
which are the generators of the R-spin operators. Their explicit forms are
given in Appendix D in Ref.\cite{Ezawa:2008ae}.

All the physical operators required for the description of the system are
constructed as the bilinear combinations of $\psi (\boldsymbol{x})$ and 
$\psi^{\dagger}(\boldsymbol{x})$. They are 16 density operators 
$\rho (\boldsymbol{x})=\psi^{\dagger}(\boldsymbol{x})\psi (\boldsymbol{x})$, 
$S_{a}(\boldsymbol{x})=\frac{1}{2}\psi^{\dagger}(\boldsymbol{x})
\tau_{a}^{\text{spin}}\psi (\boldsymbol{x})$, 
$P_{a}(\boldsymbol{x})=\frac{1}{2}\psi^{\dagger}(\boldsymbol{x})\tau_{a}^{\text{ppin}}\psi (\boldsymbol{x})$,
and $R_{ab}(\boldsymbol{x})=\frac{1}{2}\psi^{\dagger}(\boldsymbol{x})\tau
_{a}^{\text{spin}}\tau_{b}^{\text{ppin}}\psi (\boldsymbol{x})$, 
where $S_{a} $ describes the total spin, $2P_{z}$ measures the electron-density
difference between the two layers. The operator $R_{ab}$ transforms as a
spin under $\text{SU}_{\text{spin}}(2)$ and as a pseudospin under 
$\text{SU}_{\text{ppin}}(2)$. It is $R_{ab}$ that plays the key role in the entangled
spin-pseudospin phase coherence in the CAF phase.

The kinetic Hamiltonian is quenched, since the kinetic energy is common to
all states in the lowest Landau level (LLL). The Coulomb Hamiltonian is
decomposed into the SU(4)-invariant term $H_{\text{C}}^{+}$ and the
SU(4)-noninvariant term $H_{\text{C}}^{-}$. The additional potential terms
are the Zeeman, tunneling, and bias terms, $H_{\text{ZpZ}}=-\int
d^{2}x(\Delta_{\text{Z}}S_{z}+\Delta_{\text{SAS}}P_{x}+eV_{\text{bias}}P_{z})$, 
where $V_{\text{bias}}$ is the bias voltage which controls the
density imbalance between the two layers. The total Hamiltonian is 
$H=H_{\text{C}}^{+}+H_{\text{C}}^{-}+H_{\text{ZpZ}}$.

We project the density operators to the LLL. What are observed
experimentally are the classical densities, which are expectation values
such as ${\rho}^{\text{cl}}(\boldsymbol{x})
=\langle \mathfrak{S}|{\rho}(\boldsymbol{x})|\mathfrak{S}\rangle $, 
where $|\mathfrak{S}\rangle $
represents a generic state in the LLL. We may set $\rho^{\text{cl}}
(\boldsymbol{x})=\rho_{0}$, 
$S_{a}^{\text{cl}}(\boldsymbol{x})=\rho_{\Phi}\mathcal{S}_{a}(\boldsymbol{x})$,
$P_{a}^{\text{cl}}(\boldsymbol{x})=\rho_{\Phi}\mathcal{P}_{a}(\boldsymbol{x})$, 
and $R_{ab}^{\text{cl}}(\boldsymbol{x})=\rho_{\Phi}\mathcal{R}_{ab}(\boldsymbol{x})$ 
for the study of Goldstone modes, 
where $\rho_{\Phi}=\rho_{0}/\nu $ is the density of states. 
Taking the nontrivial lowest order terms in the
derivative expansion, we obtain the SU(4) effective Hamiltonian density\cite{Ezawa:2003sr} 
\begin{align}
{\mathcal{H}}^{\text{eff}}& =J_{s}^{d}\left( \sum (\partial_{k}\mathcal{S}_{a})^{2}
+(\partial_{k}\mathcal{P}_{a})^{2}+(\partial_{k}\mathcal{R}_{ab})^{2}\right)  \notag \\
& +2J_{s}^{-}\left( \sum (\partial_{k}\mathcal{S}_{a})^{2}+(\partial_{k}\mathcal{P}_{z})^{2}+(\partial_{k}\mathcal{R}_{az})^{2}\right)  \notag \\
& +\rho_{\phi}\big[\epsilon_{\text{cap}}(\mathcal{P}_{z})^{2}
-2\epsilon_{\text{X}}^{-}\left( \sum (\mathcal{S}_{a})^{2}+(\mathcal{R}_{az})^{2}\right)
\notag \\
& -(\Delta_{\text{Z}}\mathcal{S}_{z}+\Delta_{\text{SAS}}\mathcal{P}_{x}
+\Delta_{\text{bias}}\mathcal{P}_{z})\big],  \label{SU4Hamil}
\end{align}
where 
$J_{s}^{-}=\frac{1}{2}\left( J_{s}-J_{s}^{d}\right) $ with $J_{s}$ and $J_{s}^{d}$ the
intralayer and interlayer stiffness, $\epsilon_{\text{cap}}$ the
capacitance energy, $\epsilon_{\text{X}}^{-}$ the exchange Coulomb energy due to 
$H_{\text{C}}^{-}$: Their explicit formulas are given in Appendix A in Ref.\cite{Ezawa:2008ae}. 
This effective Hamiltonian is valid at $\nu =1,2,3$.

The ground state is obtained by minimizing the effective Hamiltonian (\ref{SU4Hamil}) for homogeneous configurations of the classical densities. The
order parameters are the classical densities for the ground state. They are
explicitly given in Ref.\cite{Ezawa:2005xi} for the $\nu =2$ system. In the
limit $\Delta_{\text{SAS}}\rightarrow 0$, they read 
\begin{align}
& \mathcal{S}_{z}^{0}=1-|\sigma_{0}|,\quad \mathcal{P}_{z}^{0}
=\sigma_{0},\quad \mathcal{R}_{xx}^{0}=\text{sgn}(\sigma_{0})\mathcal{R}_{yy}^{0},
\notag \\
& \mathcal{R}_{yy}^{0}=-\sqrt{|\sigma_{0}|(1-|\sigma_{0}|)},
\label{orderparameter}
\end{align}
and all others being zero. Here, $\sigma_{0}=(\rho_{0}^{\text{f}}-\rho
_{0}^{\text{b}})/(\rho_{0}^{\text{f}}+\rho_{0}^{\text{b}})$ is the
imbalance parameter with $\rho_{0}^{\text{f(b)}}$ being the electron
density in the front (back) layer. Both the spin and the pseudospin are
polarized into the $z$-axis in this limit. 

We have analyzed the excitations around the classical ground state\cite{yhama}. 
There emerge four complex Goldstone modes associated with the
spontaneous symmetry breaking SU(4)$\rightarrow $U(1)$\otimes $SU(2)$\otimes$SU(2).
When $\ H_{\text{C}}^{-}=0$ and 
$\Delta_{\text{Z}}=\Delta_{\text{SAS}}=\Delta_{\text{bias}}=0$, 
the SU(4) symmetry is exact and all  of them are gapless, but they get gapped by these interactions. 
We are interested in the limit $\Delta_{\text{SAS}}\rightarrow 0$ since we
expect the enhancement of the interlayer phase coherence just as in the 
$\nu=1$ system. We have already shown that there exists one gapless Goldstone
mode with a linear dispersion relation in a perturbation theory\cite{yhama}.

In this paper we employ the Grassmannian formalism\cite{Hasebe} to make the
physical picture of this Goldstone mode and its phase coherence clearer,
and to construct a nonperturbative theory in terms of the density difference field $\sigma (\boldsymbol{x})$
and its conjugate phase field $\vartheta (\boldsymbol{x})$. 
The Grassmannian field $Z(\boldsymbol{x})$
consists of two $\text{CP}^{3}$ fields $\boldsymbol{n}_{1}(\boldsymbol{x})$
and $\boldsymbol{n}_{2}(\boldsymbol{x})$ at $\nu =2$, since there are two
electrons per one Landau site. Due to the Pauli exclusion principle they
should be orthogonal one to another. Hence, we require 
$\boldsymbol{n}_{i}^{\dagger}(\boldsymbol{x})\cdot \boldsymbol{n}_{j}(\boldsymbol{x})
=\delta_{ij}$ 
with $i=1,2$. Using a set of two $\text{CP}^{3}$ fields  
subject to this normalization condition we introduce a $4\times 2$ matrix
field, the Grassmannian field given by 
$Z(\boldsymbol{x})=(\boldsymbol{n}_{1},\boldsymbol{n}_{2})$  
obeying $Z^{\dagger}Z=\boldsymbol{1}$.

The dimensionless SU(4) isospin densities are given by 
\begin{align}
\mathcal{S}_{a}(\boldsymbol{x})& =\frac{1}{2}\text{Tr}\left[ Z^{\dagger
}\tau_{a}^{\text{spin}}Z\right] 
=\frac{1}{2}\sum_{i=1}^{2}\boldsymbol{n}_{i}^{\dagger}\tau_{a}^{\text{spin}}\boldsymbol{n}_{i},  \notag \\
\mathcal{P}_{a}(\boldsymbol{x})& =\frac{1}{2}\text{Tr}\left[ Z^{\dagger
}\tau_{a}^{\text{ppin}}Z\right] 
=\frac{1}{2}\sum_{i=1}^{2}
\boldsymbol{n}_{i}^{\dagger}\tau_{a}^{\text{ppin}}\boldsymbol{n}_{i},
\label{su4isospin1} \\
\mathcal{R}_{ab}(\boldsymbol{x})& =\frac{1}{2}\text{Tr}\left[ Z^{\dagger
}\tau_{a}^{\text{spin}}\tau_{b}^{\text{ppin}}Z\right] =\frac{1}{2}
\sum_{i=1}^{2}\boldsymbol{n}_{i}^{\dagger}\tau_{a}^{\text{spin}}
\tau_{b}^{\text{ppin}}\boldsymbol{n}_{i},  \notag
\end{align}
where $\boldsymbol{n}_{i}$ consists of the basis 
$\boldsymbol{n}_{i}(\boldsymbol{x})=\left( n^{\text{f}\uparrow},
n^{\text{f}\downarrow},n^{\text{b}\uparrow},n^{\text{b}\downarrow}\right)^{t}$. 
It is a straightforward task to carry out the perturbative analysis of the effective
Hamiltonian (\ref{SU4Hamil}) in terms of the Grassmannian field and obtain
the same results as given in Ref.\cite{yhama}.

We concentrate solely on the gapless mode in the limit $\Delta_{\text{SAS}}\rightarrow 0$. 
We parametrize the $\text{CP}^{3}$ fields as 
\begin{equation}
\boldsymbol{n}_{1}
=\left(\begin{array}{c}
1 \\ 
0 \\ 
0 \\ 
0
\end{array}\right) ,
\quad \boldsymbol{n}_{2}
=\left(\begin{array}{c}
0 \\ 
-e^{+i\vartheta (\boldsymbol{x})/2}\sqrt{\sigma (\boldsymbol{x})} \\ 
e^{-i\vartheta (\boldsymbol{x})/2}\sqrt{1-\sigma (\boldsymbol{x})} \\ 
0
\end{array}\right) ,  \label{positivecp3}
\end{equation}
for $\sigma (\boldsymbol{x})>0$, and 
\begin{equation}
\boldsymbol{n}_{1}=
\left(\begin{array}{c}
0 \\ 
0 \\ 
1 \\ 
0
\end{array}\right) ,\quad \boldsymbol{n}_{2}=
\left(\begin{array}{c}
e^{+i\vartheta (\boldsymbol{x})/2}\sqrt{1+\sigma (\boldsymbol{x})} \\ 
0 \\ 
0 \\ 
e^{-i\vartheta (\boldsymbol{x})/2}\sqrt{-\sigma (\boldsymbol{x})}
\end{array}\right) .  \label{negativecp3}
\end{equation}
for $\sigma (\boldsymbol{x})<0$. The isospin density fields are expressed in
terms of $\sigma (\boldsymbol{x})$ and $\vartheta (\boldsymbol{x})$, 
\begin{align}
& \mathcal{S}_{z}(\boldsymbol{x})=1-|\sigma (\boldsymbol{x})|,
\quad \mathcal{P}_{z}(\boldsymbol{x})=\sigma (\boldsymbol{x}),  \notag \\
& \mathcal{R}_{yy}(\boldsymbol{x})
=\text{sgn}(\sigma_{0})\mathcal{R}_{xx}(\boldsymbol{x})
=-\sqrt{|\sigma (\boldsymbol{x})|(1-|\sigma (\boldsymbol{x})|)}\cos \vartheta (\boldsymbol{x}),  \notag \\
& \mathcal{R}_{yx}(\boldsymbol{x})=-\text{sgn}(\sigma_{0})\mathcal{R}_{xy}(\boldsymbol{x})
=-\sqrt{|\sigma (\boldsymbol{x})|(1-|\sigma (\boldsymbol{x})|)}\sin \vartheta (\boldsymbol{x}),  \label{Isospin}
\end{align}
with all others being zero. The ground-state expectation values are 
$\langle\sigma (\boldsymbol{x})\rangle =\sigma_{0}$, 
$\langle \vartheta (\boldsymbol{x})\rangle =0$, with which the order parameters 
\eqref{orderparameter} are reproduced from (\ref{Isospin}). It is notable
that the fluctuations of the phase field $\vartheta (\boldsymbol{x})$ affect
both spin and pseudospin components of the $R$-spin. This is very different
from the spin wave in the monolayer QH system or the pseudospin wave in the
bilayer QH system at $\nu =1$. Hence we call it the entangled
spin-pseudospin phase field $\vartheta (\boldsymbol{x})$.

By substituting (\ref{Isospin}) into (\ref{SU4Hamil}), apart from irrelevant
constant terms the resulting effective Hamiltonian is 
\begin{equation}
\mathcal{H}_{\text{eff}}=\frac{J_{\vartheta}}{2}\left( \nabla \vartheta
\right)^{2}+\frac{J_{\sigma}}{2}\left( \nabla {\sigma}\right)^{2}
+\rho_{\Phi}\epsilon_{\text{cap}}^{\nu =1}(\sigma -\sigma_{0})^{2},
\label{EffecHamil}
\end{equation}
where $J_{\sigma}=4J_{s}+\frac{(2|\sigma_{0}|-1)^{2}}%
{|\sigma_{0}|(1-|\sigma_{0}|)}J_{s}^{d}$, $J_{\vartheta}
=4{J_{s}^{d}}|\sigma_{0}|(1-|\sigma_{0}|)$, and $\epsilon_{\text{cap}}^{\nu =1}
=4(\epsilon_{\text{D}}^{-}-\epsilon_{\text{X}}^{-})$ is the capacitance parameter at 
$\nu =1$. 
The effective Hamiltonian is correct up to 
$\mathcal{O}(\Delta_{\text{SAS}}^{3})$ as $\Delta_{\text{SAS}}\rightarrow 0$. 

When we require the equal-time commutation relation, 
\begin{equation}
\frac{\rho_{0}}{2}\left[ \sigma (\boldsymbol{x}),\vartheta (\boldsymbol{y})\right] 
=i\delta (\boldsymbol{x}-\boldsymbol{y}),  \label{CR}
\end{equation}
the Hamiltonian (\ref{EffecHamil}) is second quantized, and it has the 
linear dispersion relation, 
\begin{equation}
E_{\boldsymbol{k}}=|\boldsymbol{k}|\sqrt{\frac{2J_{\vartheta}}{\rho_{0}}
\left( \frac{2J_{\sigma}}{\rho_{0}}\boldsymbol{k}^{2}
+2\epsilon_{\text{cap}}^{\nu =1}\right)}.
\end{equation}
This agrees with Eq.(136) of Ref.\cite{yhama}. It should be emphasized that
the effective Hamiltonian (\ref{EffecHamil}) is valid in all orders of the
phase field $\vartheta (\boldsymbol{x})$. It may be regarded as a classical
Hamiltonian as well, where (\ref{CR}) should be replaced with the
corresponding Poisson bracket.

The effective Hamiltonian (\ref{EffecHamil}) for $\vartheta (\boldsymbol{x})$
and $\sigma (\boldsymbol{x})$ reminds us of the one that governs the
Josephson-like effect at $\nu =1$. The main difference is the absence of the
tunneling term, as implies that there exists no Josephson-like tunneling. 
Nevertheless, the supercurrent is present within the layer, which is our main issue.

\begin{figure}[t]
\begin{center}
\centering
\includegraphics[width=0.48\textwidth]{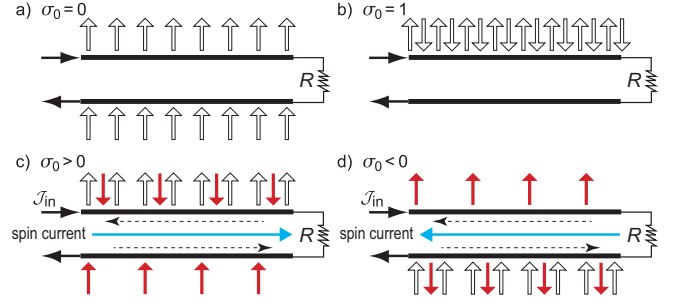}
\end{center}
\caption{(Color online) Schematic illustration of the spin supercurrent
flowing along the $x$-axis in the counterflow geometry. (a) All spins are
polarized into the positive z axis due to the Zeeman effect at $\protect\sigma_{0}=0$. 
No spin current flows. (b) All electrons belong to the front
layer at $\protect\sigma_{0}=1$. No spin current flows. (c) In the CAF
phase for $\protect\sigma_{0}>0$, some up-spin electrons are moved from the
back layer to the front layer by flipping spins. There appears a Goldstone
mode associated with this charge-spin transfer. The interlayer phase
difference $\protect\vartheta (\boldsymbol{x})$ is created by feeding a
charge current $\mathcal{J}_{\text{in}}$ to the front layer, which also
drives the spin current. Electrons flow in each layer as indicated by the
dotted horizontal arrows, and the spin current flows as indicated by the
solid horizontal arrow. (d) In the CAF phase for $\protect\sigma_{0}<0$,
similar phenomena occur but the direction of the spin current becomes
opposite.}
\label{spincurrentfigure}
\end{figure}

By using the Hamiltonian (\ref{EffecHamil}) and the commutation relation \eqref{CR}, 
we obtain the equations of motion, 
\begin{align}
\hbar \partial_{t}\vartheta (\boldsymbol{x})& 
=\frac{2J_{\sigma}}{\rho_{0}}\nabla^{2}\sigma (\boldsymbol{x})
-2\epsilon_{\text{cap}}^{\nu =1}(\sigma (\boldsymbol{x})
-\sigma_{0}),  \label{heisenbergeom1} \\
\hbar \partial_{t}\sigma (\boldsymbol{x})& =-\frac{2J_{\vartheta}}
{\rho_{0}}\nabla^{2}\vartheta (\boldsymbol{x}).  \label{heisenbergeom2}
\end{align}

\section{Anomalous Hall resistance and  Spin supercurrent} 
We now study the electric supercurrent carried by the gapless mode 
$\vartheta (\boldsymbol{x})$. The electron densities are 
$\rho_{e}^{\text{f}(\text{b})}=-{e\rho_{0}}\left( 1\pm \mathcal{P}_{z}\right) /2
=-{e\rho_{0}}\left( 1\pm \sigma (\boldsymbol{x})\right) /2$ on each layer. 
Taking the time derivative and using \eqref{heisenbergeom2} we find 
\begin{equation}
\partial_{t}\rho_{e}^{\text{f}}=-\partial_{t}\rho_{e}^{\text{b}}
=\frac{eJ_{\vartheta}}{\hbar}\nabla^{2}\vartheta (\boldsymbol{x}). 
\label{continuityequation1}
\end{equation}
The time derivative of the charge is associated with the current via the
continuity equation, $\partial_{t}\rho_{e}^{\text{f}(\text{b})}
=\partial_{i}\mathcal{J}_{i}^{\text{f}(\text{b})}$. We thus identify 
$\mathcal{J}_{i}^{\text{f(b)}}=\pm \mathcal{J}_{i}^{\text{Jos}}(\boldsymbol{x})+$constant, 
where
\begin{equation}
\mathcal{J}_{i}^{\text{Jos}}(\boldsymbol{x})\equiv \frac{eJ_{\vartheta}}{\hbar}
\partial_{i}\vartheta (\boldsymbol{x}).  \label{phasecurrent}
\end{equation}
Consequently, the current $\mathcal{J}_{i}^{\text{Jos}}(\boldsymbol{x})$
flows when there exists inhomogeneity in the phase $\vartheta (\boldsymbol{x})$. 
It is a supercurrent because the coherent mode exhibits a linear dispersion
relation. It is intriguing that the current does not flow in the balanced
system since $J_{\vartheta}=0$ at $\sigma_{0}=0$.

Let us inject the current $\mathcal{J}_{\text{in}}$ into the $x$ direction
of the bilayer sample, and assume the system to be homogeneous in the $y$
direction (Fig.\ref{spincurrentfigure}). It creates the electric field 
$E_{y}^{\text{f(b)}}$ so that the Hall current flows into the $x$-direction.
A bilayer system consists of the two layers and the volume between them. The
Coulomb energy in the volume is minimized\cite{Ezawa:2007nj} by the
condition $E_{y}^{\text{f}}=E_{y}^{\text{b}}$. We thus impose
$E_{y}^{\text{f}}=E_{y}^{\text{b}}\equiv E_{y}$. 
The current is the sum of the Hall current and the supercurrent, 
\begin{equation}
\mathcal{J}_{x}^{\text{f}}(x)=\frac{\nu}{R_{\text{K}}}
\frac{\rho_{0}^{\text{f}}}{\rho_{0}}E_{y}+\mathcal{J}_{x}^{\text{Jos}},
\quad \mathcal{J}_{x}^{\text{b}}(x)=\frac{\nu}{R_{\text{K}}}\frac{\rho_{0}^{\text{b}}}
{\rho_{0}}E_{y}-\mathcal{J}_{x}^{\text{Jos}},  \label{totalcurrent}
\end{equation}
with $R_{\text{K}}=2\pi \hbar /e^{2}$ the von Klitzing constant. We obtain
the standard Hall resistance when $\mathcal{J}_{x}^{\text{Jos}}=0$. Namely,
the emergence of the supercurrent ($\mathcal{J}_{x}^{\text{Jos}}\not=0$) is detected if the Hall
resistance becomes anomalous.

We apply these formulas to analyze the counterflow and drag experiments
since they occur without tunneling. In\ the counterflow experiment, the
current $\mathcal{J}_{\text{in}}$ is injected to the front layer and
extracted from the back layer at the same edge. Since there is no tunneling
we have 
$\mathcal{J}_{x}^{\text{b}}=-\mathcal{J}_{x}^{\text{f}}=-\mathcal{J}_{\text{in}}$. 
Hence, it follows from (\ref{totalcurrent}) that $E_{y}=0$, or
\begin{equation}
R_{xy}^{\text{f}}\equiv \frac{E_{y}^{\text{f}}}{\mathcal{J}_{x}^{\text{f}}}=0,
\qquad 
R_{xy}^{\text{b}}\equiv \frac{E_{y}^{\text{b}}}{\mathcal{J}_{x}^{\text{b}}}=0.  \label{counterflowanomalous}
\end{equation}
All the input current is carried by the supercurrent, 
$\mathcal{J}_{x}^{\text{Jos}}=\mathcal{J}_{\text{in}}$. It generates such an
inhomogeneous phase field that 
$\vartheta (\boldsymbol{x})=(\hbar/eJ_{\vartheta})\mathcal{J}_{\text{in}}x$.

On the other hand, in the drag experiment, since the interlayer coherent
tunneling is absent, no current flows on the back layer, or $\mathcal{J}_{x}^{\text{b}}=0$. 
Hence, it follows from (\ref{totalcurrent}) that 
$\mathcal{J}_{\text{in}}=\mathcal{J}_{x}^{\text{f}}=(\nu /R_{\text{K}})E_{y}$, or 
\begin{equation}
R_{xy}^{\text{f}}\equiv {\frac{E_{y}^{\text{f}}}{\mathcal{J}_{x}^{\text{f}}}=
\frac{R_{\text{K}}}{\nu}=}\frac{1}{2}R_{\text{K}}\qquad \text{at\quad}\nu=2,
 \label{draganomalous}
\end{equation}
A part of the input current is carried by the supercurrent, 
$\mathcal{J}_{x}^{\text{Jos}}=\frac{1}{2}(1-\sigma_{0})\mathcal{J}_{\text{in}}$.

The standard Hall resistance is given by 
$R_{xy}^{\text{f}}=\frac{2}{\nu}R_{\text{K}}=R_{\text{K}}$ at $\nu =2$. 
We thus predict the
anomalous Hall resistance (\ref{counterflowanomalous}) and (\ref{draganomalous}) 
in the CAF phase at $\nu =2$ by carrying out similar
experiments\cite{Kellog1,Tutuc,Kellog2} due to Kellogg \textit{et al}. and
Tutuc \textit{et al}. in imbalanced configuration ($\sigma_{0}\neq 0$).

The phase field $\vartheta (\boldsymbol{x})$ describes 
the entangled spin-pseudospin coherence according
to the basic formula (\ref{Isospin}) in the CAF phase. 
The spin density in each layer is defined by 
$\rho_{\alpha}^{\text{spin}}(\boldsymbol{x})
\equiv s_{\alpha}\psi_{\alpha}^{\dagger}\psi_{\alpha}$, 
where $s_{\alpha}=\frac{1}{2}\hbar $ 
for $\alpha =\text{f}\uparrow,\text{b}\uparrow$ and 
$s_{\alpha}=-\frac{1}{2}\hbar$ for $\alpha =\text{f}\downarrow,\text{b}\downarrow $. 
We note the relation 
\begin{equation}
\left(\begin{array}{c}
\rho_{\text{f}\uparrow}(\boldsymbol{x}) \\ 
\rho_{\text{f}\downarrow}(\boldsymbol{x}) \\ 
\rho_{\text{b}\uparrow}(\boldsymbol{x}) \\ 
\rho_{\text{b}\downarrow}(\boldsymbol{x})
\end{array}\right) =\frac{1}{4}
\left(\begin{array}{cccc}
1 & 1 & 1 & 1 \\ 
1 & -1 & 1 & -1 \\ 
1 & 1 & -1 & -1 \\ 
1 & -1 & -1 & 1
\end{array}\right) 
\left( \begin{array}{c}
\rho_{0} \\ 
2{S}_{z}(\boldsymbol{x}) \\ 
2{P}_{z}(\boldsymbol{x}) \\ 
2{R}_{zz}(\boldsymbol{x})
\end{array}\right) .
\end{equation}
Up to $\mathcal{O}((\sigma -\sigma_{0})^{2})$, we obtain 
$\mathcal{S}_{z}=1-|\sigma (\boldsymbol{x})|$, and
\begin{align}
\partial_{t}\rho_{\text{b}\uparrow}^{\text{spin}}& =\partial_{t}
\rho_{\text{f}\downarrow}^{\text{spin}}=\frac{J_{\vartheta}}{4}[1+\text{sgn}
(\sigma_{0})]\partial_{x}^{2}\vartheta (\boldsymbol{x}), \\
\partial_{t}\rho_{\text{f}\uparrow}^{\text{spin}}& =\partial_{t}
\rho_{\text{b}\downarrow}^{\text{spin}}=-\frac{J_{\vartheta}}{4}[1-\text{sgn}(\sigma_{0})]\partial_{x}^{2}\vartheta (\boldsymbol{x}).
\label{rhotderivative}
\end{align}
The time derivative of the spin is associated with the spin current via the
continuity equation, 
$\partial_{t}\rho_{\alpha}^{\text{spin}}(\boldsymbol{x})
=\partial_{x}\mathcal{J}_{\alpha}^{\text{spin}}(\boldsymbol{x})$ for
each $\alpha $. We thus identify 
\begin{align}
\mathcal{J}_{\text{b}\uparrow}^{\text{spin}}(\boldsymbol{x})& 
=\mathcal{J}_{\text{f}\downarrow}^{\text{spin}}(\boldsymbol{x})
=\frac{J_{\vartheta}}{2}\partial_{x}\vartheta (\boldsymbol{x}),\quad \text{for}\ \sigma_{0}>0, \\
\mathcal{J}_{\text{f}\uparrow}^{\text{spin}}(\boldsymbol{x})& 
=\mathcal{J}_{\text{b}\downarrow}^{\text{spin}}(\boldsymbol{x})=-\frac{J_{\vartheta}}{2}
\partial_{x}\vartheta (\boldsymbol{x}),\quad \text{for}\ \sigma_{0}<0.
\end{align}
The spin current $\mathcal{J}_{\alpha}^{\text{spin}}(\boldsymbol{x})$ flows
along the $x$-axis, when there exists an inhomogeneous phase difference $\vartheta (\boldsymbol{x})$.

In the counterflow experiment, the total charge current along the $x$-axis is zero, 
$\mathcal{J}_{x}^{\text{f}}(\boldsymbol{x})+\mathcal{J}_{x}^{\text{b}} 
(\boldsymbol{x})=0$. 
Consequently, the input current generates a pure spin current,
\begin{equation}
\mathcal{J}_{x}^{\text{spin}}=\mathcal{J}_{\text{f}\uparrow}^{\text{spin}}+
\mathcal{J}_{\text{f}\downarrow}^{\text{spin}}
+\mathcal{J}_{\text{b}\uparrow}^{\text{spin}}+\mathcal{J}_{\text{b}\downarrow}^{\text{spin}}
=\text{sgn}(\sigma_{0})\frac{\hbar}{e}\mathcal{J}_{\text{in}}.
\end{equation}
This current is dissipationless since the dispersion relation is linear. It
is appropriate to call it a spin supercurrent. It is intriguing
that the spin current flows in the opposite directions for $\sigma_{0}>0$
and $\sigma_{0}<0$, as illustrated in Fig.\ref{spincurrentfigure}. 
A comment is in order: The spin current only flows within the sample, since
spins are scattered in the resistor $R$ and spin directions become random
outside the sample.

We have explored the entangled spin-pseudospin phase coherence in the CAF
phase, governed by the Goldstone mode $\vartheta (\boldsymbol{x})$
describing the $R$-spin according to the formula (\ref{Isospin}). We have
predicted anomalous Hall resistivity in the counterflow and drag experiments
in the imbalanced regime ($\sigma_{0}\neq 0$) at $\nu =2$. In particular,
there flows a spin supercurrent in the counterflow geometry.

This research was supported in part by JSPS Research Fellowships for Young
Scientists, and a Grant-in-Aid for Scientific Research from the Ministry of
Education, Culture, Sports, Science and Technology (MEXT) of Japan (No.
21540254).

\end{document}